\documentclass[runningheads]{llncs}
\usepackage[T1]{fontenc}
\usepackage{graphicx}
\usepackage{framed}

\usepackage{hyperref} 

\begin{document}

\title{Functional Python Programming in Introductory Computer Science Courses}
\titlerunning{Functional Python Programming}

\author{Rajshekhar Sunderraman\orcidID{0000-0001-6822-6629}}
\institute{Georgia State University, Atlanta, Georgia, USA\\
\email{rsunderraman@gsu.edu}}

\maketitle              

\begin{abstract}
The functional programming paradigm has a long and storied history with its beginnings in the Lambda Calculus. In recent decades, pure functional languages such as Haskell have been shown to be highly effective in producing robust software due to immutable data structures among other functional features. The advantages of programming with immutable data structures can also be had in non-functional languages such as Python. Over the years, non-functional languages have introduced immutable data structures as well as comprehension and lambda expressions and it is possible to program in a purely functional style in them. In this paper, we present a ``best practice'' idea in introductory programming classes that forces students to learn and complete programming assignments in a purely functional subset of Python. By doing so, the student can learn functional ideas such as immutability, pure functions with no side effects, and stateless programming. We define a functional subset of Python and illustrate the best practice using small examples. We strongly feel that students in computing need familiarity with pure functional programming and argue that this can be taught in introductory programming courses that use Python. 
\end{abstract}

\section{Topic}

\vspace{-0.05in}

The functional programming paradigm has its origins in Lambda Calculus (\cite{church1932}). It was first implemented in a modern programming language, LISP (\cite{lisp}) in 1960 and during the next several decades, many functional languages were introduced either as academic learning environments (\cite{scheme}) or as industrial strength implementations (\cite{fhash}, \cite{haskell}, \cite{ocaml}).  

Functional programming offers numerous advantages in software development (\cite{whyfpmatters}). It encourages the use of pure functions (without side effects), immutability, and declarative constructs, resulting in robust, concise, and easy to understand code. At the core of functional programming is the idea of composition and decomposition. Smaller and independent functions performing a single task are constructed and composed to form the solution of a bigger problem. Re-usability of functions is a natural result of this process. Pure functions also enable testing in isolation and debugging is streamlined. With clever optimization techniques, compilers for functional programming can achieve high degree of parallelism due to the fact that there is no shared mutable state.
Immutability, a core principle of functional programming, eliminates the risk of race conditions and other concurrency issues that arise from shared mutable state. This makes it easier to write thread-safe and parallel code. Due to the fact that there is no mutable state, functional programming also reduces an entire class of common programming errors, leading to more robust and reliable software systems. With the availability of higher-order functions, a strong type system, and 
type inference, functional programming enables more expressive and high-level code. Some functional programming languages support lazy evaluation, a feature that allows for optimization and infinite data structures.

The advantages offered by the functional programming paradigm has resulted in languages in other paradigms such as Java, Python, Javascript, and Scala to incorporate many of the features of functional programming. Since most of the colleges and universities employ one or more of the non-functional languages in their introductory classes, we consider it important to teach these features as early as possible. To this end, we present an approach to incorporating functional programming in CS1 and CS2 courses. 

\section{Description of the Best Practice}

\vspace{-0.05in}

To inculcate functional programming habits in introductory computer science courses, we ask the students to complete one or more assignments using \underline{only} the following functional features of Python:

\begin{enumerate}
    
\item {\bf Restricted use of assignment statement.} The student may use variables and assign expressions to them only for the purpose of ``naming'' the expression for possible use in subsequent steps. They may also assign a new expression to a variable to discard the old value stored in the variable. Assignment statements may not be used in any other context.

\item {\bf Conditional expressions.} Expressions of the form {\tt expr1 IF cond ELSE expr2} may be used to create a value that is dependent on a condition.

\item {\bf Comprehension expressions.} Python's list, set, and dictionary comprehension expressions may be used to create a new list, set, or dictionary respectively out of existing collection objects.  

\item {\bf Lambda expressions and pure functions.} The use of lambda expressions to create small and anonymous functions for use in higher-order built-in functions is encouraged. For named functions, a strict requirement is that they are defined as ``pure'' functions that do not read or write to an external environment and produce the same output for a given input every time.

\item {\bf map, filter, and reduce.} Students may use these higher-order built-in functions to create new values from existing data. The {\tt map} and {\tt filter} built-in functions apply a specified function or predicate to elements of a collection and {\tt reduce} accumulates the results of applying a binary function on an accumulator value and an element of a collection.  

\item {\bf Immutable data structures and methods on mutable data structures that do not mutate.} Students are encouraged to use immutable data structures such a the {\tt Tuple} and {\tt String} data types in Python. They are also encouraged to use methods (such as {\tt zip()}, {\tt list()}, or {\tt sorted()}) that extract values from mutable data structures without mutating them.

\end{enumerate}

The student may not use any kind of loop-, if-, if-else-, if-elif-else- statements. Typically, the solution should contain sequence of assignment statements, where the variables on the left-hand side of the assignment is used only to ``name'' an expression. Decomposing the solution into functions is fine as long as the code within the function also follows the restrictions. Many of the local looping code can be accomplished using  comprehensions or recursive functions. 

\noindent
{\bf Example 1:} Consider encoding a string of lower case letters using Caesar Cipher method that rotates a letter by a given shift value around the alphabet. This can be solved using the following function that uses {\tt map}, {\tt reduce}, and {\tt lambda} expressions. The parameter {\tt n} is the shift value and {\tt s} is the input string.
\begin{verbatim}
def encode(n,s):
  return reduce(lambda acc,d: acc+d,
           map(lambda c: chr(ord('a')+((ord(c)-ord('a')+n)\%26))
                      if c.islower() else c,s),"")
\end{verbatim}

\noindent
{\bf Example 2:} Consider the problem of finding ``twin'' primes less than a given number, where a twin prime is a pair of primes that differ by 2. Below, we present a Python program that follows the restrictions of pure functional programming. This solution uses the classic ``sieve'' algorithm to generate all primes and then extracts consecutive primes that are twin primes.

\begin{verbatim}
def removeMultiples(x,xs):
    return [] if xs == [] else 
           removeMultiples(x,xs[1:]) if xs[0]%x == 0 else 
           [xs[0]] + removeMultiples(x,xs[1:])
def sieve(xs):
    return [] if xs == [] else 
            [xs[0]] + sieve(removeMultiples(xs[0],xs[1:]))
def filterTwins(pairs):
    return [pair for pair in pairs if pair[0]+2 == pair[1]]
def twinPrimes(n):
    ps = sieve([x for x in range(2,n+1)])
    return filterTwins(list(zip(ps,ps[1:])))
\end{verbatim}

\section{A Challenging Programming Assignment}

In this section, we present a challenging programming assignment given in a CS2 class. Students were taught the functional programming constructs discussed in this paper and were asked to produce a solution in that style for the following programming assignment.

\noindent
{\bf Programming Assignment:}
Write a Python program to compute a Relational Algebra expression corresponding to an atomic formula in Datalog. Here are two sample atomic formulas from Datalog:
\begin{verbatim}
q(x,x,x,x,x)
p(x,20,x,y,"john",x,y)
\end{verbatim}
As can be seen, an atomic formula in Datalog begins with a NAME token (denoting the relation name) followed by a left parenthesis (LPAREN), followed by a list of arguments (we will have at least one argument) and ends with a right parenthesis (RPAREN). The list of arguments is COMMA separated and each of argument can be either a NUMBER constant (assume positive integers) or a STRING constant, or a NAME corresponding to a variable.

The conversion from Datalog atom to Relational Algebra string works as follows (described in an inside-out manner, with \verb|p(x,20,x,y,"john",x,y)| as an example):

\noindent
{\bf Step 1:} Produce a list of variables, \verb|x_0, ..., x_n-1|, where n is the number of arguments in the Datalog atom and generate the following string: 
\begin{verbatim}
rename[x_0,x_1,x_2,x_3,x_4,x_5,x_6](p)
\end{verbatim}

\noindent
{\bf Step 2:} Generate the select conditions; There are two types of select conditions: one that correspond to the numeric or string constants in the Datalog atom and the other that correspond to repeated variable arguments in the Datalog atom. Combine all of the conditions with \verb|and| and generate the following string:
\begin{verbatim}
select[x_1=20 and x_4="john" and x_0=x_2 and x_2=x_5 and x_3=x_6]
  (rename[x_0,x_1,x_2,x_3,x_4,x_5,x_6](p))
\end{verbatim}
As can be seen, there are two conditions that correspond to the constant arguments of the atom, two conditions to ensure that the 3 instances of variable \verb|x| are equated and one condition to ensure that the 2 instances of the variable \verb|y| are equated. This string builds on the string generated in the previous step.

\noindent
{\bf Step 3:} Generate the project list; this list consists of the unique variables in the Datalog atom (in this case the unique variables are x and y); Generate the following string, again building on the previous string:
\begin{verbatim}
project[x_0,x_3](
  select[x_1=20 and x_4="john" and x_0=x_2 and x_2=x_5 and x_3=x_6]
    (rename[x_0,x_1,x_2,x_3,x_4,x_5,x_6](p)))
\end{verbatim}

\noindent
{\bf Step 4:} Produce the rename list of unique variables from the original names of variables and generate the final string:
\begin{verbatim}
rename[x,y](project[x_0,x_3](
  select[x_1=20 and x_4="john" and x_0=x_2 and x_2=x_5 and x_3=x_6]
    (rename[x_0,x_1,x_2,x_3,x_4,x_5,x_6](p))))
\end{verbatim}

\noindent
Here are two functions to develop to solve the problem:

\begin{verbatim}
# Input is the Datalog atom. data = "p(x,20,x,y,"john",x,y)"
# Returns pair (predicate, arguments), predicate = "p" and
#   arguments = [('var','x'),('num',20),('var','x'),('var','y'),
                 ('str','John'),('var','x'),('var','y')]
def parseF(data):
    pass

# Input is same as output of parseF
# Returns Relational Algebra expression for input
def convert2ra(rname_args):
    pass
\end{verbatim}

\section{Justification}

\vspace{-0.05in}

While it is agreed upon that writing functional style code is a good thing, unfortunately most curriculum in computing start with imperative programming and focus more on ``state-based'' programming with emphasis on maintaining state in variables and processing the data by writing loops and conditional statements. The best practice introduced in this paper can be incorporated into the imperative style of teaching by making the students write pure functions and use higher-level expressions to transform data. By asking the students to program in this style in introductory classes can help them in forming a habit and it is likely that in the future the student will write functional style code in their work.

\section{Outcomes and Insights}

\vspace{-0.05in}

The best practice has been used in a CS2 course in the computer science curriculum and anecdotally the programming assignment using the functional style was perceived to be difficult but exciting by the students. The assignment generated the most ``buzz'' in the class with much more interaction in the class discussion board as well as more students showing up for office hours to get clarifications. The expectation is that the students who have been exposed to functional style in early classes will turn out to be better programmers. 

\vspace{-0.05in}

In this paper, we have presented an approach to introduce functional programming in Python in introductory programming courses. The ability to use features such as immutable data, pure functions, higher-order functions, and functional expressions in writing code is an essential skill for every Computer Scientist. By learning to code in a purely functional style in Python, students can produce robust solutions to problems in their subsequent courses and in their careers. 

Although many curriculum use Python as the first programming language due to its simplicity and its use in Data Science, it is important for computer scientists to learn strongly-typed languages in subsequent courses in the curriculum. A suggestion to fellow educators who may adopt the best practice idea is to repeatedly emphasize to the student the importance of strongly-typed languages to ensure robustness of code.

\newpage

\bibliographystyle{splncs04}
\bibliography{Main}

\end{document}